\documentclass[preprint, showpacs,preprintnumbers,amsmath,amssymb,showkeys]{revtex4}
\usepackage{graphicx}

\begin{document}
\title{Electromagnetic and Scalar Pion form factor in the Kroll-Lee-Zumino model}
\author{C. A. Dominguez$^{\mathrm{a,b}}$,
        J. I. Jottar$^{\mathrm{c,d}}$,
        M. Loewe$^{\mathrm{c}}$,
        B. Willers$^{\mathrm{a}}$\\
$^{\mathrm{a}}$Centre for Theoretical Physics and Astrophysics,
University of
Cape Town\\ Rondebosch 7700, South Africa \\
$^{\mathrm{b}}$Department of Physics,
Stellenbosch University\\ Stellenbosch 7600, South Africa\\
    $^{\mathrm{c}}$Pontificia Universidad Cat\'{o}lica de Chile,\\ Casilla
306, Santiago 22, Chile\\
     $^{\mathrm{d}}$Department of Physics,
University of Illinois, \\Urbana-Champaign,  IL, 61801 3080, USA}

\begin{abstract}
\noindent
The renormalizable Abelian quantum field theory model of
Kroll, Lee, and Zumino is used at the one loop level to compute
vertex corrections to the tree-level, Vector Meson Dominance
 (VMD) electromagnetic pion form factor. These corrections, together with the one-loop vacuum
  polarization contribution, imply a resulting electromagnetic pion form
  factor in excellent agreement with data in the whole range
   of accessible momentum transfers in the space-like region. The time-like form factor, which
   reproduces
    the Gounaris-Sakurai formula at and near the rho-meson peak, is unaffected
    by the vertex correction at order $\cal{O}$$(g^2)$. The KLZ
    model is also used to compute the scalar radius of the pion at
    the one loop level, finding $\langle r^{2}_{\pi }\rangle_{S} =
    0.40 fm^{2}$. This value implies for the low energy constant of
    chiral perturbation theory $\bar{l}_{4} =3.4$.

\end{abstract}
\maketitle
\noindent
 The renormalizable Abelian quantum field theory of charged
pions, and  massive neutral
 vector mesons, proposed by Kroll, Lee, and Zumino (KLZ) \cite{KLZ}, provides
  a rigorous theoretical justification for the Vector Meson Dominance (VMD) ansatz \cite{VMD}. Since
  the  neutral vector mesons are coupled
   only to conserved currents the model is renormalizable \cite{KLZ},\cite{Hees}.
   Gale and  Kapusta  \cite{GK} computed in this model
the rho-meson self energy to one-loop order. When this result is
used in the VMD expression for the electromagnetic pion form factor,
the Gounaris-Sakurai formula \cite{GS}-\cite{tau} in the time-like
region is found at and near the rho-meson pole. This is quite
intriguing. That an empirical fit formula such as this should follow
from  the KLZ Lagrangian may be hinting at additional unexpected
properties of this model. Here we compute the vertex diagram, i.e.
the one loop correction to the strong coupling constant in the
framework of the KLZ model. This correction is of the same order in
the coupling as the one loop vacuum polarization. After going
through regularization and renormalization, and in conjunction with
the VMD expression for the electromagnetic pion form factor, this
vertex correction, together with the vacuum polarization
contribution, leads to an excellent agreement between theory and
data in the spacelike region. The parameter free result (masses and
couplings are known from experiment) represents a substantial
improvement over naive (tree-level) VMD. The resulting chi-squared
per degree of freedom is close to unity, while the one from
tree-level VMD is about five times bigger. Time-like region
predictions are unaffected by the vertex correction. In fact, the
combination of vacuum polarization and vertex corrections in this
region turns out to be of higher order in the coupling. Since the
KLZ model involves a strong coupling, the perturbative expansion
could be questioned, and the next-to-leading (one-loop)
contributions need not be smaller than the leading term. However,
this is not the case in the KLZ model. In fact, the relatively small
$\rho\pi\pi$ coupling ($g_{\rho\pi\pi} \simeq 5$) is accompanied by
the large loop suppression factor $1/(4 \pi)^2$, so that the
one-loop contributions remain reasonable corrections to the leading
order tree-level term. At higher orders, we expect higher powers of
this suppression factor from loop integrations. A
next-to-next-to leading order calculation is beyond the scope of this work.\\

 \noindent The KLZ lagrangian is given by

\begin{eqnarray}
L_{KLZ}  =  \partial_\mu \phi \; \partial^\mu \phi^* -\frac{1}{4}
m_{\pi }^{2} \;\phi \;\phi^* - \frac{1}{4}\; \rho_{\mu \nu }
\;\rho^{\mu \nu} \nonumber\\ + \frac{1}{2}\; m_\rho ^{2}\; \rho_{\mu
} \;\rho^{\mu } +g_{\rho\pi \pi } \rho_{\mu} J^{\mu}_{\pi} \;,
\end{eqnarray}

\noindent
 where $\rho_\mu$ is a vector field describing the $\rho^0$
meson ($\partial_\mu \rho^\mu = 0$), $\phi$ is a complex
pseudo-scalar field describing the $\pi^\pm$ mesons, $\rho_{\mu\nu}$
is the usual field strength tensor, and $J^\mu_\pi$ is the $\pi^\pm$
current, i.e.
\begin{equation}
\rho_{\mu \nu}  = \partial_\mu \rho_\nu - \partial_\nu \rho_\mu \; ,
\end{equation}

\begin{equation}
J^{\mu }_{\pi }  =  i (\phi^{*}\partial ^{\mu }\phi - (\partial
^{\mu }\phi ^{*})\phi ) \; .
\end{equation}

\noindent It is important to remark that the KLZ lagrangian includes
also a coupling of the form $g_{\rho\pi \pi}^{2} \rho _{\mu }
\rho^{\mu }  \phi \phi ^{*}$ which we omitted from Eq.1, since it is
not relevant for our calculation.

\noindent In Fig. 1 we show the vertex function kinematics. The
double line denotes the $\rho $-propagator and the dashed lines the
pion propagators.
\begin{figure}
\includegraphics[scale=0.4]{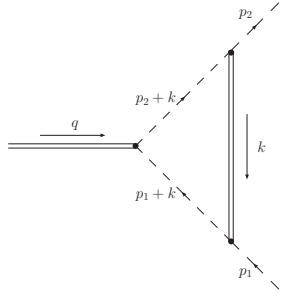}
\caption{Vertex function kinematics} \label{fig1}
\end{figure}
The technical details associated to the evaluation of this diagram
can be found in the original reference \cite{DJLW}. Here we will
only present the main results. In the Feynman gauge and using
dimensional regularizarion, \cite{Hees},\cite{Quigg} our vertex
function can be written as
\begin{eqnarray}
\widetilde{\Gamma} ^{(1)\mu}_{\rho \pi \pi}({p_1}, {p_2},
q^2)=\Gamma ^{(0)\mu}_{\rho \pi \pi}({p_1},
{p_2})[\widetilde{G}(q^{2})+
\nonumber\\
 A ( \frac{2}{\varepsilon} -\frac{1}{2} - \gamma +
\ln (4 \pi) ) ] + \mathcal{O}(\varepsilon)
\end{eqnarray}
\noindent where the tree-level contribution is given by
\begin{equation}
\Gamma^{(0)\mu}_{\rho \pi \pi} (p_1, p_2) = i g_{\rho \pi \pi}
\mu^{(2 - \frac{d}{2})}(p_1+p_2)^\mu \;,
\end{equation}
and
\begin{eqnarray}
&& \hspace{-.5cm} \widetilde{G}(q^2) =
 - 2 \frac{g^2_{\rho \pi\pi}}{(4 \pi)^2}
\int_{0}^{1}\! dx_{1}\!\int_{0}^{1-x_{1}}
 \!\!\!\!\!dx_{2} \Big\{(2 - 3x_1) \nonumber\\
 && \hspace{-0.2cm} \ln \left( \frac{\Delta(q^2)}{\mu^2} \right)
 + \left( \frac{1 - 2 x_1}{2 \Delta(q^2)} \right)\Big{[} m_\pi^2 (x_1 + x_2 - 2)^2
\nonumber\\
&& - q^2(x_1 x_2 - x_1 - x_2 + 2) \Big{]} \Big\} \;. \label{Gsquig}
\end{eqnarray}

\noindent The factor $A$ is a certain integral over $x_{1}$ and
$x_{2}$ but it does not depend on $q^{2}$. So, this constant will be
canceled during the renormalization procedure. In the previous
equations $\mu $ is the scale mass factor associated to dimensional
regularization and

\begin{equation}
\Delta (q^{2}) = m_{\pi }^{2}(x_{1} +x_{2})^{2}+ m_{\rho
}^{2}(1-x_{1}-x_{2})-x_{1}x_{2}q^{2}
\end{equation}

\noindent It is convenient to choose $q^{2}=0$ as the
renormalization point for the vertex function, because we can make
use of the well known normalization of the pion form factor $F_{\pi
}(0) = 1$,
 together
with

\begin{equation}
\Gamma^{(1)\mu}_{\rho\pi\pi} (p_1,p_2,q^2 = 0)  =
\Gamma^{(0)}_{\rho\pi\pi}(p_1,p_2) \;.
\end{equation}

\noindent The electromagnetic pion form factor in VMD at tree level
is given by
\begin{equation}
F_{\pi}(q^2)|_{\mbox{VMD}} = \frac{g_{\rho\pi\pi}}{f_{\rho}}\;
\frac{M_{\rho}^2}{M_{\rho}^2 - q^{2} }\;.
\end{equation}

\noindent The renormalized electromagnetic pion form factor
associated to the substracted one-loop vertex corrections at
$q^{2}=0$, at order $\cal{O}$$(g_{\rho\pi\pi}^2)$, can be written as

\begin{eqnarray}
&&F_{\pi}(q^2)|_{\mbox{vertex}} =
F_{\pi}(q^2)|_{\mbox{VMD}}\nonumber\\
&&\times [1+ \widetilde{G}(q^2) - \widetilde{G}(0)],
\end{eqnarray}

\noindent where $f_\rho = 4.97 \pm 0.07$ \cite{PDG}, and from
universality and $F_\pi(0) = 1$ it follows that $g_{\rho\pi\pi}(0) =
f_\rho$. Hence, the one-loop vertex correction generates an
additional momentum dependence in the form factor.

\medskip
\noindent Two seagull-type corrections of the same order in
$\cal{O}$$(g_{\rho\pi\pi}^2)$in principle should be included.
However these diagrams do not depend on $q^{2}$ and they cancel when
substracting at $q^{2}=0$. Another contribution to the pion form
factor comes from the vacuum polarization corrections to the $\rho
$-propagator. This calculation has been done in \cite{GK}. See
\cite{DJLW} for details. Adding the vacuum polarization to the
vertex contribution gives the complete correction to the VMD
electromagnetic pion form factor at order
$\cal{O}$$(g_{\rho\pi\pi}^{2})$. We emphasize that the result does
not contain free parameters, since the masses and the coupling are
known from experiment.

\begin{eqnarray}
&&F_\pi(q^2) = \frac{M_\rho^2 + \Pi(0)|_{\mbox{vac}}}{M_\rho^2 - q^2
+ \Pi(q^2)|_{\mbox{vac}}} \nonumber\\
&&+ \frac{M_\rho^2}{M_\rho^2 - q^2} \Big[ \widetilde{G}(q^2) -
\widetilde{G}(0)\Big] \;,
\end{eqnarray}

\noindent where
\begin{eqnarray} &&\Pi(q^2)|_{\mbox{vac}} =
\frac{1}{3}\; \frac{g_{\rho\pi\pi}^{2}}{(4 \pi)^2} \; \;q^2 \;
\;\Big(1 - 4\; \frac{\mu_\pi^2}{q^2}\Big)^{3/2} \nonumber\\
 && \times \left[ \ln \Bigg| \frac{\sqrt{(1 - 4 \;\mu_{\pi}^{2}/q^2)} + 1}{\sqrt{(1 - 4 \;
 \mu _{\pi}^{2}/q^2)} - 1} \Bigg| \right. \nonumber \\
&& \left. -i \; \pi\; \theta(q^2 - 4 \mu_\pi^2)
\phantom{\frac{1}{1}} \right] + A \; q^2 + B \; ,
\end{eqnarray}
and where the constants $A$ and $B$ are given by

\begin{eqnarray}
&&A = - \frac{1}{3} \; \frac{g_{\rho\pi\pi}^2}{(4 \pi)^2}\; \Bigg{[}
8\; \frac{\mu_{\pi}^2}{M_{\rho}^2} + \Big{(}1 -  4 \;
\frac{\mu_{\pi}^2}{M_{\rho}^2}\Big{)}^{3/2}  \nonumber\\
&& \times  \ln \Bigg| \frac{\sqrt{(1 - 4 \;\mu_{\pi}^2/M_{\rho}^2)}
+ 1}{\sqrt{(1 - 4\; \mu_{\pi}^2/M_{\rho}^2)} - 1} \Bigg| \Bigg{]}
\;,
\end{eqnarray}

\begin{equation}
B = \Pi(0)|_{\mbox{vac}} = \frac{8}{3} \; \frac{g_{\rho\pi\pi}^2}{(4
\pi)^2}\; \mu_\pi^2 \;.
\end{equation}

\noindent It is important to remark that the vacuum polarization
correction is not included in the second term of Eq.11 as it would
make this term of $\cal{O}$$(g^4)$. Hence, the vertex correction
does not affect the form factor in the time-like region, where it
becomes the Gounaris-Sakurai formula near the rho-meson peak. In
fact, from the definition of the hadronic width \cite{PIL}:
$\Gamma_\rho = - (1/M_\rho)\;\,Im \; \Pi(M_\rho^2)  $, where
$\Gamma_\rho \equiv \Gamma_\rho(M_\rho^2)$, and from Eq.(12) there
follows

\begin{equation}
\Gamma_\rho = \frac{g_{\rho\pi\pi}^2}{48 \pi}\; \frac{1}{M_\rho^2}
\; (M_\rho^2 - 4 \, \mu_\pi^2)^{\frac{3}{2}} \;,
\end{equation}

\noindent which is the standard kinematical relation between width
and coupling of a vector and two pseudoscalar particles \cite{PIL}.
Notice that this results follows automatically in the KLZ model,
i.e. it has not been imposed as a constraint. Near the rho-meson
peak, where $\Pi(s)$ is largely purely imaginary,  the s-dependent
width which follows from Eqs. (12) and (15) is

\begin{equation}
\Gamma_\rho(s)|_{KLZ} = \frac{M_\rho\, \Gamma_\rho}{\sqrt{s}}
\Big[\frac{s - 4\, \mu_\pi^2}{M_\rho^2 - 4\,
\mu_\pi^2}\Big]^{\frac{3}{2}} \;,
\end{equation}

\noindent which is precisely the momentum dependent Gounaris-Sakurai
width \cite{PIL}.
This is known to provide an excellent fit to the data in this region \cite{tau}. \\

\noindent Our results are shown in Fig.2 (solid line) together with
the experimental data \cite{data} and the prediction from tree-level
VMD (dotted curve). The latter provides a poor fit to the data as
evidenced from the resulting chi-square per
 degrees of freedom $\chi_F^2 = 5.0$, while Eq.(11) gives the optimal value $\chi_F^2 = 1.1$.
 In addition, the mean-square radius of the pion obtained from Eq.(35) is $<r^2_\pi> = 0.46 \;\mbox{fm}^2$,
 to be compared with a much smaller result from tree-level VMD $<r^2_\pi> = 6/M_\rho^2 = 0.39 \;\mbox{fm}^2$, and
the experimental value
  $<r^2_\pi> = 0.439 \; \pm \;0.008\; \mbox{fm}^2$.

\begin{figure}[ht]
\begin{center}
\includegraphics[width=\columnwidth]{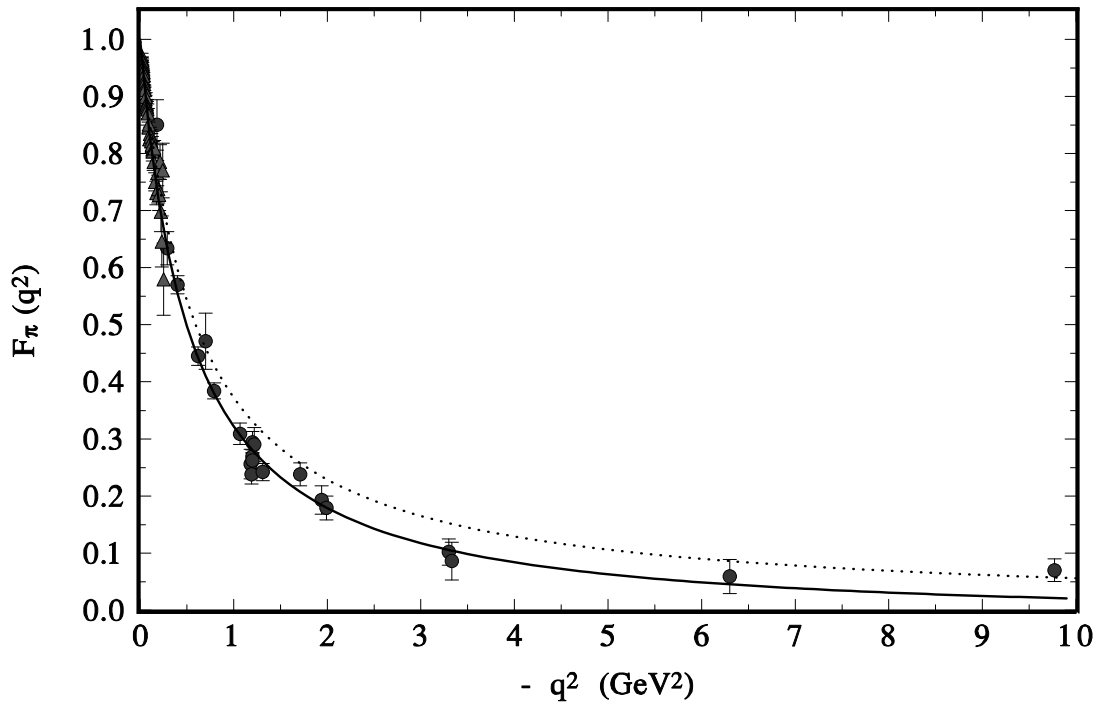}
\caption{Pion form factor data together with the KLZ prediction,
Eq.(11) (solid line), and the tree-level  VMD result (dotted line).}
\end{center}
\end{figure}


  \noindent In summary, the KLZ one-loop level contributions to the pion form factor turn out to be reasonable
corrections to the leading order result. This is in spite of KLZ
being a strong interaction theory. This is due to the relatively
mild coupling ($g_{\rho\pi\pi} \simeq 5$), together with a large
loop suppression factor ($(1/4 \pi)^2)$, as seen from Eq. (6).
Increasing powers of this suppression factor are expected at higher
orders in perturbation theory.

\medskip
\noindent In the mean time, the KLZ model has been used to calculate
the scalar radius of the pion at next to leading order (one loop) in
perturbation theory. See \cite{DLW} for details. The result provides
additional support for the KLZ theory as a viable platform to
compute corrections to VMD systematically in perturbation theory.

\noindent The pion scalar form factor $\Gamma _{\pi }(q^{2})$ is
given in terms of the pion matrix elements of the scalar operator
$J_{S} = m_{u} \bar{u}u + m_{d}\bar{d}d$ according to

\begin{equation}
\Gamma _{\pi }(q^{2}) = \langle \pi (p_{2})|J_{S}|\pi (p_{1})\rangle
,
\end{equation}
\noindent where $q^{2} = (p_{2}-p_{1})^{2}$. The associated
quadratic scalar radius

\begin{equation}
\Gamma _{\pi }(q^{2}) = \Gamma _{\pi }(0)\left[1 + \frac{1}{6}
\langle r_{\pi }^{2}\rangle _{S}q^{2} + \cdots\right],
\end{equation}
is important in chiral perturbation theory since it fixes
$\bar{l}_{4}$ \cite{CHPT}.

\smallskip
\noindent Our one loop corrections in the KLZ model give
\begin{equation}
\langle r_{\pi }^{2}\rangle _{S} = 0.4 fm^{2}; \; \bar{l}_{4} = 3.4,
\end{equation}
\noindent which are close to the current values \cite{oller}
$\langle r_{\pi }^{2}\rangle _{s} \simeq  0.5 - 0.7fm^{2}$ and
$\bar{l}_{4} \simeq 4.0-5.1$. Higher loop corrections $(NNLO)$ are
currently been analized, since they are the main uncertainty in this
determination.


\bigskip
\noindent
 {\bf Acknowledgements:}   We acknowledge support by Fondecyt
 under grants 1051067, 7070178, and by Centro de Estudios
 Subat\'omicos (Chile) and by NRF (South Africa)

\end{document}